\documentstyle[12pt,psfig]{article}

\def\baselinestretch{1.2}
\begin{document}

\begin{titlepage}

\begin{flushright}
OHSTPY-HEP-T-02-003 \\
hep-th/0202071
\end{flushright}
%\vspace{12 mm}
\vfil\vfil

\begin{center}

{\Large {\bf Numerical Simulations of $N=(1,1)$ SYM$_{1+1}$ with 
Large Supersymmetry Breaking}}

\vfil
I.~Filippov and S.~Pinsky \\
\vfil
Department of Physics \\ Ohio State University, Columbus, OH~~43210\\
\end{center}

\begin{abstract}
We consider the $N=(1,1)$ SYM theory that is obtained by dimensionally 
reducing SYM theory in 2+1 dimensions to 1+1 dimensions and discuss
soft supersymmetry breaking.  We discuss the numerical simulation of 
this theory using SDLCQ when either the boson or the fermion has a 
large mass. We compare our result to the pure adjoint 
fermion theory and pure adjoint boson DLCQ calculations of Klebanov, 
Demeterfi, and Bhanot and of Kutasov. With a large boson mass we find 
that it is necessary to add additional operators to the theory to obtain 
sensible results. When a large fermion mass is added to the theory we 
find that it is not necessary to add operators to obtain a sensible theory. 
The theory of the adjoint boson is a theory that has stringy bound states 
similar to the full SYM theory. We also discuss another theory of adjoint 
bosons with a spectrum similar to that obtained by Klebanov, Demeterfi,
and Bhanot.
\end{abstract}

%\vspace{20mm}
\vfil\vfil\vfil
\begin{flushleft}
January 2002
\end{flushleft}

\end{titlepage}
\renewcommand{\baselinestretch}{1.05}  %Line spacing

\section{Introduction}

In recent years we have worked extensively on a numerical method~\cite{mss95} 
for solving exactly supersymmetric field theories in $1+1$ and $2+1$ dimensions. We call this method
supersymmetric discrete light-cone quantization (SDLCQ), and we have successfully
applied it to many theories and addressed a number of interesting
issues~\cite{alp,appt98,alppt98,ahlp99,hlpt00,hhlp99,hpt01,fph01}. 
The world is, however, not exactly supersymmetric, and it is therefore 
important to learn how to generalize SDLCQ to solve non-supersymmetric theories. 

With this objective in mind we focus in this work on the interrelation of the
numerical simulations of four separate theories. Two of these theories are
supersymmetric, and two are not. The objective is to learn something about soft
supersymmetry breaking within the context of these numerical simulations. We want
to know if softly broken theories make sense in the context of SDLCQ and how
these broken theories are related to non-supersymmetric theories of adjoint
fermions and  adjoint bosons that have been discussed in the literature. 

The two non-supersymmetric simulations are those of  Klebanov, Demeterfi and
Bhanot~\cite{dkb93,dkb94} (KDB). The first is a calculation of a gauged adjoint
fermion in 1+1 dimensions and the second is a calculation of a gauged adjoint
boson in $1+1$ dimensions. The third theory we will consider is again a gauged
adjoint fermion theory in 1+1 dimensions but now with a special mass value
that makes the theory supersymmetric~\cite{Kutasov93}. Finally, the fourth 
theory is the
$N=(1,1)$ SYM that one obtains by dimensionally reducing SYM in
2+1 dimensions to 1+1 dimensions~\cite{mss95,alp}. Henceforth we will refer
to this theory as SYM.  Starting with this fourth theory we consider the 
theories that are obtained by adding a large mass for one of the fields. 
The large mass freezes out that field, leaving the low-mass bound states to  
have primarily only constituents of the other field. 

The numerical method that we use to simulate these theories is discrete light-cone
quantization (DLCQ). When this method is applied to the continuum hamiltonian of
a theory, it is called just DLCQ, and it produces a finite dimensional
hamiltonian. When this method is applied to the continuum supercharge of a
supersymmetric theory, it produces a finite dimensional supercharge which is then
used to calculate a finite dimensional hamiltonian, $P^- = (Q^-)^2/\sqrt{2}$. We
refer to this approximation as supersymmetric DLCQ or SDLCQ. 

To discretize the
hamiltonian or supercharge, we introduce discrete longitudinal momenta $k^+$ as
fractions $nP^+/K$ of the total longitudinal momentum $P^+$, where $K$ is an
integer that determines the resolution of the discretization and is known in DLCQ
as the harmonic resolution~~\cite{bpp98}. We then convert the
mass eigenvalue problem $2P^+P^-|M\rangle = M^2 |M\rangle$\, ($P^\pm=(P^z\pm
P^0)/\sqrt{2}$) to a matrix eigenvalue problem by introducing a basis where
$P^+$ is diagonal.  Because light-cone longitudinal momenta
are always positive, $K$ and each $n$ are positive integers; the number of
constituents is then bounded by $K$.  The continuum limit is recovered by
taking the limit $K \rightarrow \infty$. 

Of course, we can write the continuum hamiltonian for a supersymmetric theory and
apply DLCQ to it. This yields a different finite dimensional approximation to the
hamiltonian than does SDLCQ. Recently we have developed a technique for writing
down directly finite-dimensional DLCQ hamiltonians that are identical to the
hamiltonians obtained in SDLCQ~\cite{fph01}.

In constructing the discrete approximation we drop the longitudinal zero-momentum
mode.  For some discussion of dynamical and constrained zero modes, see the
review~\cite{bpp98}  and previous work~\cite{alp}. Inclusion of these modes
would be ideal, but the techniques required to include them in a numerical
calculation have proved to be difficult to develop, particularly because of
nonlinearities.   For DLCQ calculations that can be  compared with exact
solutions, the exclusion of zero modes does not affect the massive
spectrum~\cite{bpp98}. In scalar theories it has been known for some time that 
constrained zero modes can give rise to dynamical symmetry breaking~\cite{bpp98}.

In section 2 we discuss SYM in both the SDLCQ and
DLCQ formulation. We consider in some detail the singular terms in the hamiltonian
formulation and their action on the Fock basis. In section 3 we  add a
large boson mass to this theory and discuss the resulting pure
fermion theory. We compare our results to the fermionic theories of KDB and
Kutasov~\cite{dkb93,dkb94,Kutasov93}.  In section 4 we add a large fermionic mass
to the full SYM theory and discuss the pure boson theory that is obtained.
We find a new theory of adjoint bosons with properties similar to the full SYM
theory.  We consider the behavior of this theory with a small bare mass term 
for the boson and the relation of this theory to the KDB theory of adjoint 
bosons.  We also present yet another theory similar to the KDB theory that 
arises from the discussion. In section 5 we summarize the results of this investigation of supersymmetry breaking and
the implication for higher dimensional theories.

\section{The Supersymmetric Theory}

The SYM is given by the supercharge~\cite{mss95}
%
%supercharge definition
\begin{eqnarray}
\label{Qminus}
Q^-&=& {{\rm i} 2^{-1/4} g \over \sqrt{\pi}}\int_0^\infty dk_1dk_2dk_3
\delta(k_1+k_2-k_3) \left\{ \frac{}{} \right.\nonumber\\
&&{1 \over 2\sqrt{k_1 k_2}} {k_2-k_1 \over k_3}
[a_{ik}^\dagger(k_1) a_{kj}^\dagger(k_2) b_{ij}(k_3)
-b_{ij}^\dagger(k_3)a_{ik}(k_1) a_{kj}(k_2) ]\nonumber\\
&&{1 \over 2\sqrt{k_1 k_3}} {k_1+k_3 \over k_2}
[a_{ik}^\dagger(k_3) a_{kj}(k_1) b_{ij}(k_2)
-a_{ik}^\dagger(k_1) b_{kj}^\dagger(k_2)a_{ij}(k_3) ]\nonumber\\
&&{1 \over 2\sqrt{k_2 k_3}} {k_2+k_3 \over k_1}
[b_{ik}^\dagger(k_1) a_{kj}^\dagger(k_2) a_{ij}(k_3)
-a_{ij}^\dagger(k_3)b_{ik}(k_1) a_{kj}(k_2) ]\nonumber\\
&& ({ 1\over k_1}+{1 \over k_2}-{1\over k_3})
[b_{ik}^\dagger(k_1) b_{kj}^\dagger(k_2) b_{ij}(k_3)
+b_{ij}^\dagger(k_3) b_{ik}(k_1) b_{kj}(k_2)]  \left. \frac{}{}\right\}\,,
\end{eqnarray}
which in turn defines the hamiltonian by virtue of the anticommutation
relation $\{Q^-,Q^-\} = 2 \sqrt{2} P^-$. Throughout this paper we will write
expressions in a continuum form for notational convenience; however, it is to be
understood that all the calculation are discrete in momentum space. The numerical
method SDLCQ simply means that we apply DLCQ to the supercharge and then square
the finite dimensional representation of the supercharge to get the Hamiltonian. 
Recently we found the hamiltonian that in the DLCQ approximation reproduces the 
SDLCQ hamiltonian
%
%hamiltionian
\begin{eqnarray}
P^-&=&\frac{g^2N_c}{4\pi}\int_0^\infty dk_1 \frac {\mu^2(k_1)}{k_1}
(a^\dagger a +b^\dagger b)+\frac{g^2}{4\pi}\int_0^\infty dk_1dk_2dk_3dk_4[
\nonumber\\ &+&
   A_1 b^\dagger b^\dagger b b+
   A_2 (b^\dagger b b b - b^\dagger b^\dagger b^\dagger b) +
   B_1 a^\dagger a^\dagger a a +
	B_2 (a^\dagger a a a + a^\dagger a^\dagger a^\dagger a)
\nonumber\\ & + &
	C_1 b^\dagger b^\dagger a a + C_2 a^\dagger a^\dagger b b +
   C_3 b^\dagger a^\dagger b a + C_4 a^\dagger b^\dagger a b +
   C_5 b^\dagger a^\dagger a b + C_6 a^\dagger b^\dagger b a
   \nonumber \\ &+&
   D_1 (a^\dagger a b b - a^\dagger b^\dagger b^\dagger a) +
   D_2 (a^\dagger b a b - b^\dagger a^\dagger b^\dagger a) +
   D_3 (a^\dagger b b a - b^\dagger b^\dagger a^\dagger a)
\nonumber\\ &+&
   D_4 (b^\dagger b a a + b^\dagger a^\dagger a^\dagger b) +
   D_5 (b^\dagger a b a + a^\dagger b^\dagger a^\dagger b) +
   D_6 (b^\dagger a a b + a^\dagger a^\dagger b^\dagger b)]\,,
   \end{eqnarray}
where the coefficient in front of the dynamic mass term is given by~\cite{fph01}
\begin{equation}
\mu^2(k_1)=\int_0^{k_1} dk_2\frac{{(k_1+k_2)}^2}{k_2{(k_1-k_2)}^2}=
\int_0^{k_1} dk_2 (\frac{4k_1}{(k_2-k_1)^2} + \frac{1}{k_2}).
\end{equation}
It is convenient to define the instantaneous mass contributions to the
Hamiltonian, $P^-_{I{\rm mass}}({\rm boson})$ and 
$P^-_{I{\rm mass}}({\rm fermion})$, as
\begin{eqnarray}
P^-_{I{\rm mass}}({\rm boson})&=&\frac{g^2N_c}{\pi}\int_0^\infty dk_1a(k_1)^\dagger a(k_1)
\int_0^{k_1}  \frac{dk_2}{(k_2-k_1)^2}\,,
\nonumber\\
P^-_{I{\rm mass}}({\rm fermion})&=&\frac{g^2N_c}{\pi}\int_0^\infty dk_1b(k_1)^\dagger b(k_1)
\int_0^{k_1}  \frac{dk_2}{(k_2-k_1)^2}\,,
\end{eqnarray}
which are part of $\mu$ defined above.  The coefficients of the pure fermion and
the pure boson terms are
\begin{eqnarray}
A_1 &=& PV\frac{2}{(k_4-k_2)^2}-\frac{2}{(k_1+k_2)^2} -
\delta_{1,3}(\frac{2}{k_1^2}+\frac{2}{k_2^2})\,,
\nonumber\\
A_2 &=& \frac{2}{(k_2+k_3)^2}-\frac{2}{(k_1+k_2)^2}\,,
\nonumber\\
B_1 &=& \frac{1}{\sqrt{4k_1k_2k_3k_4}}
\left(\frac{(k_1-k_2)(k_3-k_4)}{(k_1+k_2)^2}-
PV\frac{(k_1+k_3)(k_2+k_4)}{(k_4-k_2)^2}\right)\,,
\nonumber\\
B_2 &=& \frac{1}{\sqrt{4k_1k_2k_3k_4}}
\left(\frac{(k_3-k_2)(k_1+k_4)}{(k_3+k_2)^2}+
\frac{(k_1-k_2)(k_3+k_4)}{(k_1+k_2)^2}\right)\,.
\end{eqnarray}
The term $\delta_{1,3}$ only makes sense in a discrete formulation and vanishes 
in the continuum limit; however, it does not appear in a conventional DLCQ 
formulation. This term arises from normal ordering the square of the discrete 
supercharges rather than normal ordering the continuum formulation of the 
supercharge and then discretizing this
result. These terms are part of what insures that this hamiltonian is exactly
supersymmetric. A detailed discussion of the origin of these term can be found 
in~\cite{fph01}. Similarly, when obtaining the SDLCQ hamiltonian version of the
Kutasov model one finds for the same reason a different discrete formula for 
the mass term than in conventional DLCQ . 

The singularities in $P_{I{\rm mass}}$ cancel the singularities in $A_1$ and
$B_1$. This cancellation is commonly seen in light-cone calculations and is not 
related to supersymmetry. We  demonstrated that this occurs for both the pure 
fermionic and pure bosonic theories by considering the effect of these terms on
Fock states. We will see that KDB in their theory of adjoint bosons treat these 
singularities differently than they are treated in SDLCQ and in the KDB and 
Kutasov treatment of adjoint fermions.

The $1/k_2$ term in $\mu$ is a real logarithmic mass divergence.
In a non-supersymmetric theory this requires a mass renormalization. Here in a
supersymmetric theory the bound states are such that this term is finite.  

It is interesting to compare this to the QCD$_2$ model
described by Kutasov~\cite{Kutasov93}, where
%kutasov supercharge definition
\begin{eqnarray}
\label{QKutasov}
Q^-&=& {{\rm i} 2^{-1/4} g \over \sqrt{\pi}}\int_0^\infty dk_1dk_2dk_3
\delta(k_1+k_2-k_3) 
({ 1\over k_1}+{1 \over k_2}-{1\over k_3})
[b_{ik}^\dagger(k_1) b_{kj}^\dagger(k_2) b_{ij}(k_3) \nonumber\\ &+& 
b_{ij}^\dagger(k_3) b_{ik}(k_1) b_{kj}(k_2)].  
\end{eqnarray}
From normal ordering the discrete supercharge we find that the hamiltonian 
for this theory contains the terms $A_1$, $A_2$, 
$P^-_{I{\rm mass}}({\rm fermion})$ and the mass
term with $m^2=\frac{g^2N_c}{\pi}$. We note that it does not contain the real
logarithmic divergence seen in SYM.

%%%%%%%%%%%%%%%%%%%%%%%%%%%%%%%%%%%%%%%%%%%%%%%%%%%%%%%%%%%%%%%%%%%%%%%%%%%%%%%
\section{Pure Adjoint Fermion Theories}
%%%%%%%%%%%%%%%%%%%%%%%%%%%%%%%%%%%%%%%%%%%%%%%%%%%%%%%%%%%%%%%%%%%%%%%%%%%%%%%
The pure adjoint fermion theories that we wish to consider are the DLCQ
hamiltonian theory studied by KDB~\cite{dkb93}, the supersymmetric model of
Kutasov~\cite{Kutasov93}, and the pure fermion theory one obtains by including a
large mass for the adjoint bosons in SYM. 
Let us start by considering the effect of $A_1$ and $P^-_{I{\rm mass}}$ 
on a typical Fock state
\begin{equation}
|bbb>= \int_0^\infty \frac{ds dn dk}{N^{3/2}} f(s,n,k) {\rm Tr}[b^\dagger(s) b^\dagger(n)
b^\dagger(k) ]|0>.
\end{equation}
We find
\begin{eqnarray}
P^-_{I{\rm mass}}|bbb>&=&\frac{3g^2N}{2\pi}\int^\infty_0
\frac{dsdndk}{N^{3/2}} \delta(s+n+k-P^+){\rm Tr}(b^\dagger(s)b^\dagger(n)b^\dagger(k)|0>
\nonumber \\
&&\times\int^{s+n}_0\frac{dt}{(s-t)^2}f(s,n,k)\,. 
\end{eqnarray}
The factor of three is from using the cyclic symmetry to combine the action on all
three permutations. The effect of the singular term in $A_1$ on this Fock state is 
\begin{eqnarray}
P^-_{\rm sing}|bbb>&=&\frac{-3g^2N}{2\pi}\int^\infty_0
\frac{dsdndk}{N^{3/2}} \delta(s+n+k-P^+){\rm Tr}(b^\dagger(s)b^\dagger(n)b^\dagger(k)|0>
\nonumber \\
&&\times\int^{s+n}_0\frac{dt}{(s-t)^2}f(t,s+n-t,k)\,. 
\end{eqnarray}
Again the factor of three is from using the cyclic symmetry to combine the action
on all three permutations. We see that the singularities in these terms at
$s=t$ cancel, leaving an integral that is well defined and finite in the 
principal value prescription. In Ref.~\cite{dkb93} the authors explicitly 
consider the integral equation for the Hamiltonian of adjoint fermions and find 
this cancellation.  Numerically all the theories involving adjoint fermions handle 
this singularity the same way. They include both terms and omit the discrete 
point $s=t$. 

It is already known that the KDB~\cite{dkb93} and Kutasov~\cite{Kutasov93}
formulations produce the same numerical results for a bare mass squared of 
$g^2N_c/\pi$ in the continuum limit.  We want to consider the SYM theory 
and add a larger adjoint boson mass term ($x=m^2_{\rm boson} \pi/g^2 N$) to 
the square of the supercharge, which will freeze out the bosons. Of course,
if we also subtract from the square of the supercharge the logarithmically
divergent fermion mass term and add a bare mass term with
$m^2=g^2N_c/\pi$, we reproduce the results of the Kutasov model. For 
completeness we have checked this, and it works exactly for bound states well 
below the mass of the heavy boson.

\begin{figure}
\begin{tabular}{cc}
\psfig{figure=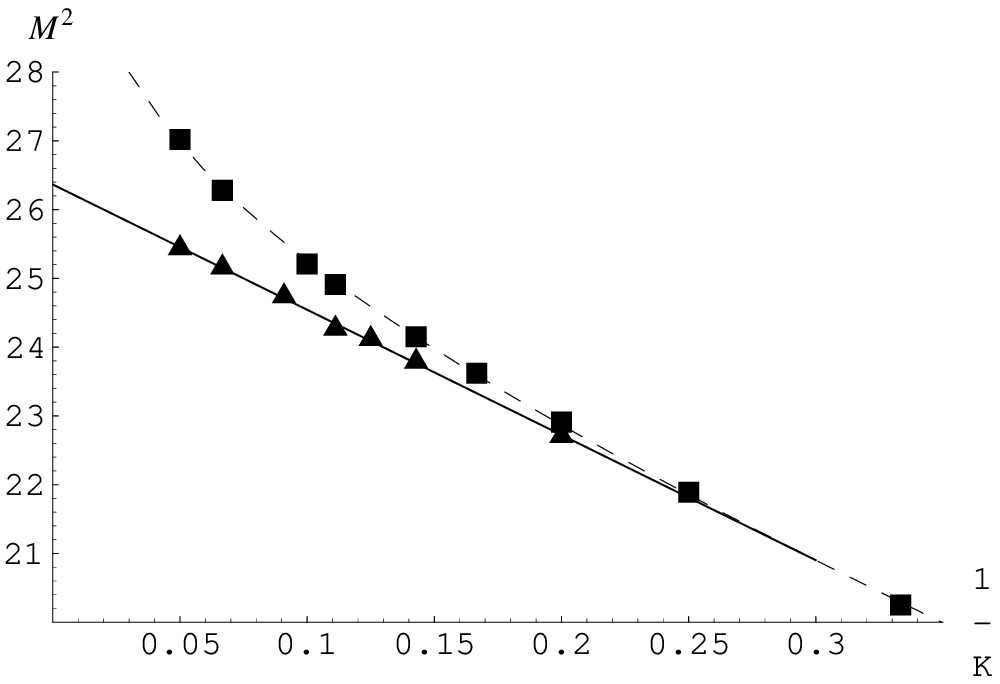,width=7.5cm,angle=0} &
\psfig{figure=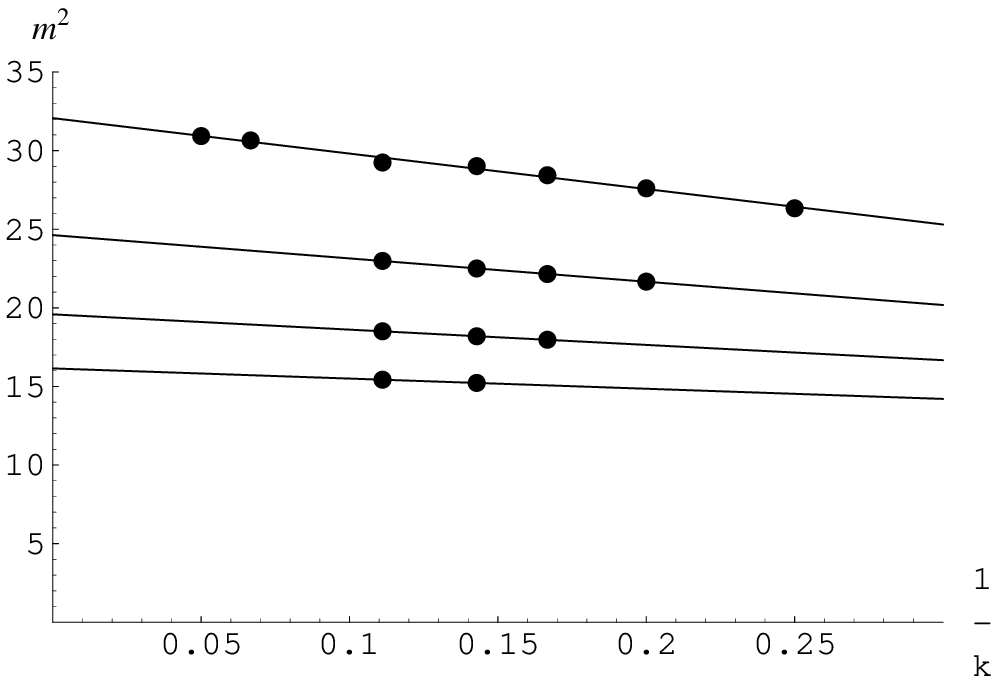,width=7.5cm,angle=0}\\
(a) & (b)
\end{tabular}
\caption{ ${\pi M^2\over g^2 N_c}$ vs ${1 \over K}$, where $K$ is the
resolution, for SYM with (a) a large boson mass and (b) a very large 
fermion mass.  In (a), the triangles represent the Kutasov model,
for comparison.}
\label{fig1}
\end{figure}

The main question that we consider in this section is whether the theory we
obtain by simply adding a large mass for the boson is a sensible theory. We 
have looked at this numerically and the answer is no! It produces 
negative-mass bound states that do not disappear as we increase the resolution. 
We can repair this problem by adding a bare mass for the adjoint fermion, but 
this does not guarantee a sensible theory. For example, if we add a mass of
$m^2=g^2N_c/\pi$, we effectively have the Kutasov model with a logarithmically
divergent mass term included. We expect this model to produce bound-state masses
that diverge logarithmically. In Fig.~\ref{fig1}(a) we have plotted the lowest
bound states of this theory and of the Kutasov model.
The fit to the mass of this theory is
\begin{equation}
M^2=\frac{g^2 N_c}{\pi}[23.22 - 13.66 \frac{1}{K} - 1.48 \log(K)]\,.
\end{equation}
Therefore the pure fermionic theories that are obtained by adding a
large boson mass require a mass renormalization exactly as we expected.  

Within the context of SDLCQ we have also looked at the decoupling of the 
heavy bosonic states in more detail. At resolution $K=4$, for example, we can 
look at the evolution of the lowest-mass bound states as a function of the boson mass $x$ in units of $g^2N_c/\pi$.  At this resolution, the pure fermionic state
$b^\dagger (3) b^\dagger (1)$
mixes with  $a^\dagger (3) a^\dagger (1)$ and  $a^\dagger (2) a^\dagger (2)$.
The corresponding hamiltonian matrix for the set of states
\{$b^\dagger (3) b^\dagger (1)$, $a^\dagger (3) a^\dagger (1)$,
$a^\dagger (2) a^\dagger (2)$\} is
\begin{displaymath}
\left( \begin{array}{ccc}
21.88 & -0.96  & 1.41 \\
-0.96 & 8.33 + 5.33x & -12.24 \\
1.41 & -12.24 & 18+4x
\end{array} \right)\,.
\end{displaymath}
At large $x$ the lowest mass eigenstate is $\{1,\theta_1,\theta_2\}$, where
$\theta_1$ and $\theta_2$ are the bosonic components. We find that $\theta_1$ 
and $\theta_2$ fall off linearly with $1/x$. 
%%%%%%%%%%%%%%%%%%%%%%%%%%%%%%%%%%%%%%%%%%%%%%%%%%%%%%%
% End of insert. (Also see the graph below.
%%%%%%%%%%%%%%%%%%%%%%%%%%%%%%%%%%%%%%%%%%%%%%%%%%%%%%%
%
%
%%%%%%%%%%%%%%%%%%%%%%%%%%%%%%%%%%%%%%%
%%%%%%%%%%%%%%%%%%%%%%%%%%%%%%%%%%%%%%%%%%%%%%%%%%%%%%%%%%%%%%%%%%%%%%%%%%%%%%%%%
\section{Pure Adjoint Boson Theories}
%%%%%%%%%%%%%%%%%%%%%%%%%%%%%%%%%%%%%%%%%%%%%%%%%%%%%%%%%%%%%%%%%%%%%%%%%%%%%%%%%
The pure boson theories that we want to compare are the theory of
KDB and SYM theory with a large
mass for the adjoint fermion. Physically the bosons in 
these theories are the transverse gluons of
the (2+1)-dimensional parent theory; therefore, these states are effectively 
(1+1)-dimensional glueballs. 

First we consider the result of simply adding a large fermion mass to the
SYM theory. This leaves us with the pure boson theory which contains the 
$P^-_{I{\rm mass}}$ term and the logarithmically divergent mass term, just 
as in the fermion theory. Recall that the fermion
theory with the logarithmically divergent mass had a divergent spectrum, but surprisingly the low-mass states of the boson theory, seen in Fig.~\ref{fig1}(b), are linear as functions of
$1/K$, and therefore the continuum spectrum obtained by extrapolating to
$K=\infty$ appears to remain finite. We have calculated one state out to
resolution $K=20$, to check for any logarithmic dependence, and found none. 
A partial explanation for this has to lie in the stringy nature of these bound 
states. In the spectrum of the pure SYM we found that, as we increased the 
resolution, new bound states appeared with more partons and with a lower mass. 
This abundance of low-mass states with many
partons is what we refer to as the stringy nature of the theory. Interestingly 
we see this property for this pure glue theory here in Fig.~\ref{fig1}(b). 
The fact that the logarithmically divergent mass term does not give rise to a divergent spectrum as a function of the
resolution is apparently related to the stingy nature of the bound states.  
 
We now want to contrast this stringy theory with the model considered by
KDB~\cite{dkb94}. There are two main differences between these two models. 
First, KDB renormalizes away the logarithmically divergent mass term.   
Second, they use a different approach to treat the singularity in the
$P^-_{I{\rm mass}}$ term and the singular term in $B_1$. This issue is a little
complicated, so let us start our discussion by considering the effect of these 
two terms on a typical Fock state
\begin{equation}
|aaa>= \int_0^\infty \frac{ds dn dk}{N^{3/2}} f(s,n,k) 
{\rm Tr}[a^\dagger(s) a^\dagger(n) a^\dagger(k) ]|0>\,.
\end{equation}
We find
\begin{eqnarray}
P^-_{I{\rm mass}}({\rm boson})|aaa>&=&\frac{3 g^2N}{2\pi}\int^\infty_0
\frac{dsdndk}{N^{3/2}} \delta(s+n+k-P^+)
     {\rm Tr}(a^\dagger(s)a^\dagger(n)a^\dagger(k)|0>
\nonumber \\
&&\times \int^{s+n}_0\frac{dt}{(n-t)^2}f(s,n,k)\,. 
\label{Imassb}
\end{eqnarray}
This is, of course, the same as the result for adjoint fermions. The singular
interaction term in $B_1$ gives
\begin{eqnarray}
P^-_{\rm sing}|aaa>&=& \frac{-3g^2N}{4\pi}\int^\infty_0
\frac{dsdstdk}{N^{3/2}} \delta(s+n+k-P^+)
    {\rm Tr}(a^\dagger(s)a^\dagger(n)a^\dagger(k)|0>
\nonumber \\
&&\times\int^{s+n}_0dt\frac{(t+n)(n+2s-t)f(t,s+n-t,k)}
                      {2\sqrt{nst(n+s-t)}\,(n-t)^2} \,.
\label{singb}
\end{eqnarray}
This expression has a complicated coefficient not present for fermions.  
We see that at the singular point $n=t$ the two terms cancel; therefore, the 
combination is non-singular in the principal value prescription. Numerically, 
in the SDLCQ calculation, these two terms are simply included in their 
discrete form with the singular points removed. This exactly parallels the 
adjoint fermion calculations discussed above.

KDB treat the singularities differently, however. To regularize the
singularity in $B_1$ they add and subtract a term 
\begin{eqnarray}
&&\frac{-3g^2N}{4\pi}\int^\infty_0
\frac{dsdstdk}{N^{3/2}} \delta(s+n+k-P^+){\rm Tr}(a^\dagger(s)a^\dagger(n)a^\dagger(k)|0>
\nonumber \\
&&\times f(s,n,k)\int^{s+n}_0dt
    \frac{(t+n)(n+2s-t)}{2\sqrt{nst(n+s-t)}\,(n-t)^2} \,.
\label{convergence}
\end{eqnarray}
The part that is subtracted makes Eq.~(\ref{singb})
finite in the principal value sense. For the part that is added they cut out the
singular point and do the remaining principal value integral exactly and include
it in the integral equation. The singular part cancels exactly the singularity in
$P_{I{\rm mass}}$.

The remaining finite part of $P_{I{\rm mass}}$ has the form of a mass term with 
$m^2 = -2 g^2 N/\pi$. They lump this term together with the logarithmically divergent mass term
and a bare mass term, to form a renormalized mass term. In principle this
appears to be simply a different way of renormalizing the mass and a different 
way of making the singularity finite in the principal value prescription. 

Numerically KDB find a very different spectrum than we found above.
Their spectrum is very QCD-like, with the lowest-mass bound states having
primarily two gluons and masses of about $4 g^2 N_c/\pi$. The higher mass states
have dominant components with a small number of particles.
KDB did the calculation using anti-periodic boundary conditions, and we have
repeated the calculation using both periodic and anti-periodic boundary
conditions. We get the same results by both methods, and our anti-periodic
calculation agrees exactly with KDB. The convergence is similar in both methods
when one takes into account the fact the you have to go to twice the
resolution with anti-periodic boundary conditions to get the same number of data
points. 

It is interesting to make a direct comparison of the KDB approach with
SDLCQ. To make this comparison we will repeat the SDLCQ calculation, but now we
will drop the divergent mass term and add a bare mass term as KDB do and 
then calculate the spectrum for various values of the bare mass. 
%%%%%%%%%%%%%%%%%%%%%%%%%%%%%%%%%%%%%%%%%%%%%%%%%%%%%%%
\begin{figure}
\begin{tabular}{cc}
\psfig{figure=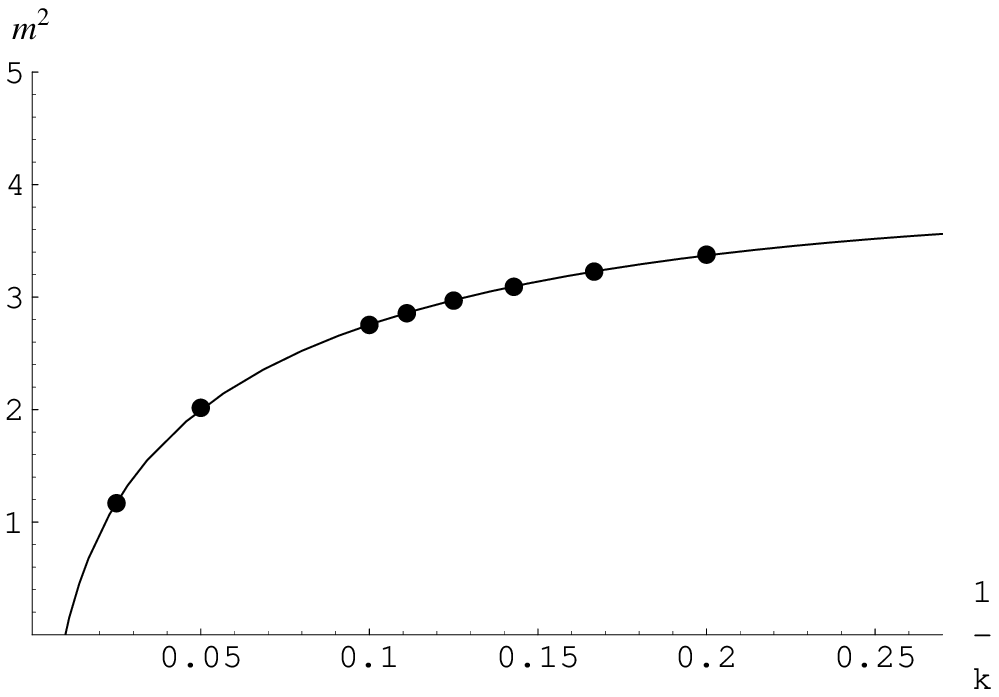,width=7.5cm,angle=0} &
\psfig{figure=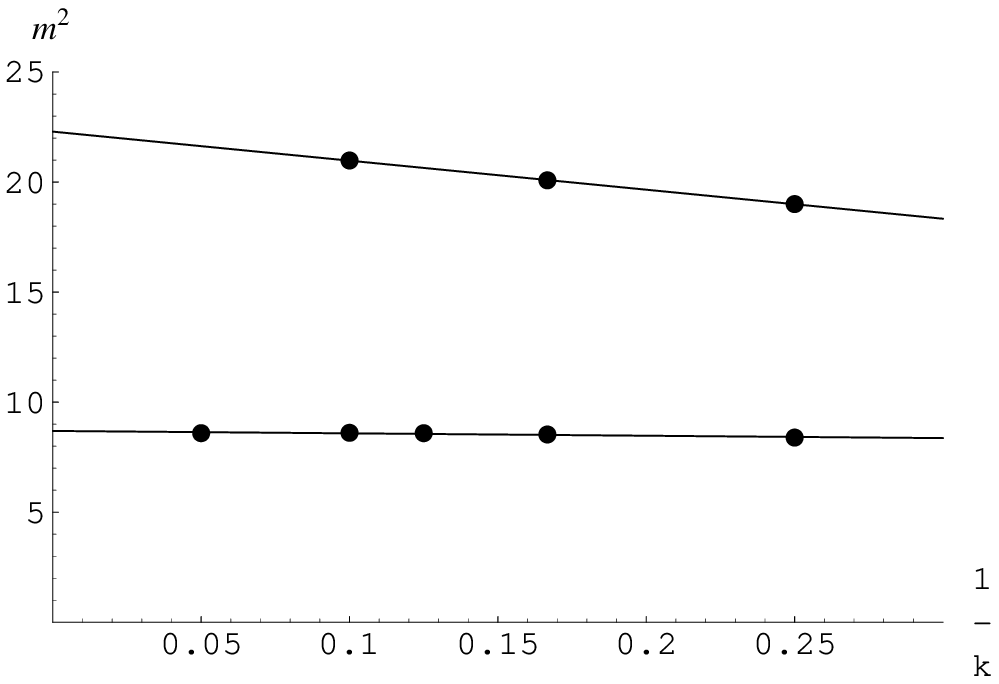,width=7.5cm,angle=0} \\
(a) & (b) 
\end{tabular}
\caption{ ${\pi M^2\over g^2 N_c}$ vs ${1 \over K}$, where $K$ is the resolution,
for the SDLCQ solution of SYM 
with addition of a heavy fermion mass, subtraction of the logarithmically
divergent boson mass, and addition of a bare mass of (a) $m^2=g^2 N_c/\pi$ or 
(b) $m^2=2g^2 N_c/\pi$.}
\label{fig2}
\end{figure}
%%%%%%%%%%%%%%%%%%%%%%%%%%%%%%%%%%%%%%%%%%%%%%%%%%%%%%%
With the bare mass equal to zero the theory is unstable and generates a negative
mass for the bound states. For $m^2=g^2N_c/\pi$ the mass of the lowest bound
state as a function of the resolution is shown in Fig.~\ref{fig2}(a). It clearly
does not converge well as a function of $1/K$. It is possible that in the
continuum limit the mass of this state goes to zero, but it also possible that 
it becomes negative. The fit shown is
\begin{equation}
M^2=\frac{g^2N_c}{\pi}[6.00 - 2.80\frac{1}{K} + 1.29078 \log(K)].
\end{equation}
The lowest-mass bound state for $m^2=2 g^2N_c/\pi$ is shown in
Fig.~\ref{fig2}(b), and we see that the bound state is well behaved and 
is fit nicely by a linear plot in
$1/K$. In fact the entire spectrum for this theory is well behaved. An 
inspection of the wave functions of this theory shows that it has a valence structure similar to KDB but not identical.

%%%%%%%%%%%%%%%%%%%%%%%%%%%%%%%%%%%%%%%%%%
\section{Discussion}
%%%%%%%%%%%%%%%%%%%%%%%%%%%%%%%%%%%%%%%%%%
We have considered the effect of adding a large boson or fermion mass to SYM and 
compared the results with those known for pure fermionic and pure bosonic 
theories discussed in the literature~\cite{dkb93,dkb94,Kutasov93}. 
We find that when we add a large boson mass the
SYM theory the resulting theory is not a sensible theory unless we also subtract a
logarithmically divergent mass term.  Otherwise this logarithmically divergent 
mass term causes the spectrum to diverge logarithmically with the resolution. 
When we subtract the logarithmically divergent mass (renormalize it away), 
we find the one-parameter family of theories considered by KDB and Kutasov. 

When we add a large fermion mass to the SYM theory, the resulting bosonic theory
does appear to be a sensible theory. The masses are linear in $1/K$ and have a 
well defined continuum limit. The spectrum is very stringy, as is the spectrum of 
the original SYM theory.  As we increase the resolution, we find states with lower 
masses and more partons.  In comparing it to the work of KDB, we see that they 
renormalized away the logarithmically divergent mass term, treat the Coulomb 
singularity differently from the way it is treated in SDLCQ, and add a bare mass.
The spectrum that they 
find also converges well in $1/K$, and the spectrum is very QCD-like. The low-mass 
states have a few valence partons, and, as one increases the resolution, one finds 
higher mass states. When we renormalize away the logarithmically divergent mass 
and add a bare mass but use the SDLCQ treatment of the Coulomb singularity, 
we also find a QCD-like spectrum for some values of the bare mass.  If the bare
mass is too small, however, we find that the spectrum is badly behaved. 

In summary, adding a large fermion mass to SDLCQ produces a sensible theory 
while adding a large boson mass does not. The SDLCQ hamiltonian formulation in 
1+1 dimensions is simple enough that we are able to identify the operator 
that needed to be added or altered to produce sensible theories.  In higher 
dimensions there is a larger list of operators to consider and this will 
represent a significant challenge.

%%%%%%%%%%%%%%%%%%%%%%%%%%%%%%%%%%%%%%%%%%
\section{Acknowledgments}
This work was supported in part by the US Department of Energy. One of the authors
(S.P) would like to acknowledge the hospitality of the Aspen Center of Physics
where part of the work was completed. The authors would like to acknowledge
conversations with U. Trittmann, O. Lunin, and J. Hiller.

\end{document}